\documentclass[conference]{IEEEtran}
\IEEEoverridecommandlockouts
\usepackage{amsmath, amssymb, amsfonts}
\usepackage{graphicx}
\usepackage[hidelinks]{hyperref}
\usepackage{url}
\usepackage{textcomp}
\usepackage{xcolor}
\usepackage{siunitx}
\usepackage{comment}
\usepackage{tabularx}
\usepackage{longtable} % tables that can span several pages
\usepackage{colortbl}
\usepackage{multirow, multicol, array, makecell}
\usepackage{booktabs, stfloats}
\usepackage{subcaption}
\usepackage{cleveref}
\usepackage{algorithmic}
\usepackage{enumitem}
\usepackage{bm}
\usepackage{flushend}
\usepackage{glossaries}

\newacronym{rwd}{RWD}{Random Window Discriminator}
\newacronym{tts}{TTS}{text-to-speech}
\newacronym{asv}{ASV}{Automatic Speaker Verification}
\newacronym{mae}{MAE}{Mean Absolute Error}
\newacronym{mse}{MSE}{Mean Squared Error}
\newacronym{auc}{AUC}{Area Under the Curve}
\newacronym{eer}{EER}{Equal Error Rate}
\newacronym{tnr}{TNR}{True Negative Rate}
\newacronym{tpr}{TPR}{True Positive Rate}
\newacronym{fnr}{FNR}{False Negative Rate}
\newacronym{roc}{ROC}{Receiver Operating Characteristic}
\newacronym{fad}{FAD}{Fréchet Audio Distance}

\begin{document}

\title{Detector-Guided Vocoder Training for \\ Adversarial Text-to-Speech Synthesis

% \thanks{This work was supported by the FOSTERER project, funded by the Italian Ministry of Education, University, and Research within the PRIN 2022 program.
% This work was partially supported by the European Union - Next Generation EU under the Italian National Recovery and Resilience Plan (NRRP), Mission 4, Component 2, Investment 1.3, CUP D43C22003080001, partnership on ``Telecommunications of the Future’’ (PE00000001 - program ``RESTART’’) and ``SEcurity and RIghts in the CyberSpace’’ (PE00000014 - program ``FF4ALL-SERICS’’).}
}

\author{\IEEEauthorblockN{
Jiayan Cui,
Davide Salvi,
Paolo Bestagini,
Stefano Tubaro
}\vspace{0.35em}
\IEEEauthorblockA{\textit{Dipartimento di Elettronica, Informazione e Bioingegneria (DEIB), Politecnico di Milano}\\
%Piazza Leonardo Da Vinci 32, 20133 Milano, Italy\\
jiayan.cui@mail.polimi.it, \{davide.salvi, paolo.bestagini, stefano.tubaro\}@polimi.it
}
}

\maketitle

\begin{abstract}
In this paper, we propose a novel Text-to-Speech (TTS) framework designed to perform adversarial attacks against speech deepfake detection systems.
Given a TTS model and a detector capable of classifying its outputs as synthetic, our approach fine-tunes the vocoder by integrating the detector into the TTS training loop.
Unlike traditional adversarial methods that rely on post-hoc perturbations or external noise injection, our approach enables the generation of synthetic speech that is intrinsically adversarial, capable of lowering the target detector's reliability without requiring additional post-processing steps.
We evaluate the framework in both white-box and black-box settings, showing its effectiveness against known and unseen detectors. 
We also propose a multi-detector strategy that allows the system to simultaneously learn to deceive multiple detectors, showcasing the scalability of the method.
Experimental results confirm the efficacy of our approach, which has the potential to impact forensic security and detection countermeasures.
At the same time, subjective listening tests reveal that the proposed method also enhances synthesized speech quality.
\end{abstract}


\section{Introduction}
\label{sec:introduction}

Recent advances in speech synthesis and deep learning have led to significant progress in \gls{tts} technologies, enabling the generation of highly realistic speech with minimal human effort~\cite{oord2016wavenet, wang2017tacotron, ren2020fastspeech}.
These technologies have become increasingly important for applications such as accessibility, entertainment, education, and human-computer interaction. At the same time, however, they have also facilitated the creation of speech deepfakes, i.e., synthetic speech signals that closely resemble authentic human speech and can impersonate a target speaker or convey arbitrary spoken content.
The increasing realism of these systems has raised significant concerns regarding misinformation, fraud, identity theft, and other malicious applications, motivating extensive research on reliable speech deepfake detection. 

To address these threats, the multimedia forensics community has developed a wide range of speech deepfake detectors.
Most of these rely on supervised learning, ranging from the use of traditional signal processing techniques~\cite{albadawy2019detecting, xue2022audio}, to that of more sophisticated models, such as Graph Attention Networks~\cite{jung2022aasist} and Transformers~\cite{bartusiak2023transformer, cuccovillo2024audio}.
Other strategies exploit transfer learning from pre-trained systems~\cite{li2023voice, yang2024robust}, or analyze semantic inconsistencies in the signals~\cite{conti2022deepfake, attorresi2022combining} to determine their authenticity.

In parallel with the development of forensic detection tools, there has been growing interest in anti-forensic techniques. These aim to identify and exploit vulnerabilities in detection systems by designing adversarial attacks~\cite{xie2020real, neekhara2019universal}.
In the audio domain, adversarial attacks have predominantly been implemented as post-processing operations.
These involve injecting learned additive noise or perturbations into audio signals to hinder their forensic traces and deceive the detectors that are used to process them~\cite{carlini2018audio, liu2019adversarial, zhang2021attack, wang2025bim}.
For instance, \cite{panariello2023malafide} proposed an attack that injects convolutional noise using a linear time-invariant filter, while~\cite{todisco2024malacopula} developed a noise injection attack based on a generalized Hammerstein model to target specific speech deepfake detectors.
Similarly, \cite{zhang2022waveform} presented a joint adversarial attack capable of deceiving both \gls{asv} systems and speech deepfake detectors simultaneously, while \cite{li2024initial} proposed a novel attack in a physical access context, where a replay process is involved.
Despite their efficacy, the dependence on post-processing is a key limitation of these methods.
Adversarial processing remains a separate stage of the generation pipeline, introducing additional computational complexity and potentially leaving artifacts that can affect perceptual quality or be exploited by more robust forensic methods.

In this paper, we depart from the paradigm of additive noise-based attacks and propose a novel approach that integrates adversarial perturbation directly into the speech synthesis process.
Specifically, we fine-tune the vocoder of a \gls{tts} system by incorporating the loss of one or more speech deepfake detectors into the vocoder training objective.
During optimization, the detector provides differentiable feedback that guides the vocoder towards generating speech signals that retain high perceptual quality while being classified as authentic by the target detector. As a result, the generated speech is intrinsically adversarial and does not require any additional waveform perturbation or post-processing at inference time. 

The primary contributions of this work are the following:
\begin{itemize}[leftmargin=*]
   \item \emph{Detector-guided adversarial synthesis} – We propose a detector-guided vocoder optimization framework that generates intrinsically adversarial speech without requiring post-processing operations.
   \item \emph{Real-time applicability} – Since adversarial fine-tuning occurs offline, the system generates adversarial speech in real time without additional runtime overhead.
   \item \emph{Generalization capability} - We evaluate the method under white-box and black-box conditions, showing that it effectively deceives both targeted and unseen detectors. Furthermore, we introduce a multi-detector strategy, enabling simultaneous attacks on multiple architectures.
   \item \emph{Enhanced speech quality} – Beyond adversarial robustness, our fine-tuning improves perceptual quality, as validated through a subjective listening test.
\end{itemize}

\section{Proposed Method}
\label{sec:method}

\begin{figure}
\centering
\includegraphics[width=.6\columnwidth]{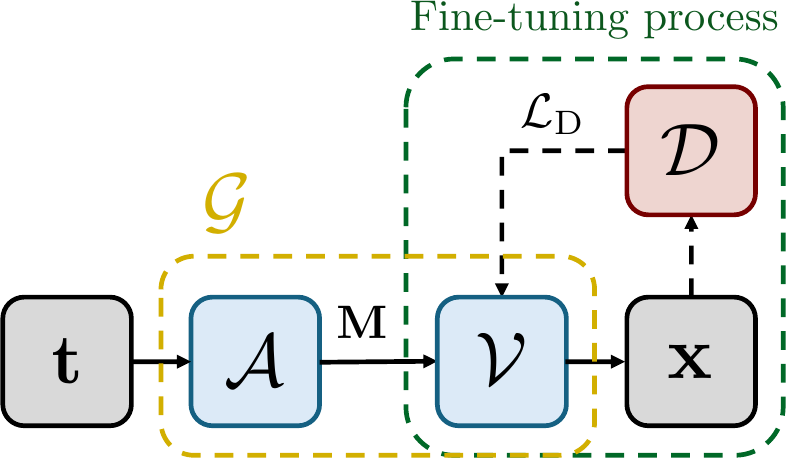}
\caption{Pipeline of the proposed \gls{tts} adversarial model.}
\label{fig:tts-pipeline}
% \vspace{-0.75em}
\end{figure}

In this paper, we propose an adversarial \gls{tts} framework that generates speech intrinsically optimized to evade speech deepfake detectors.
Rather than applying adversarial perturbations after synthesis, our approach incorporates the detector directly into the vocoder optimization process, enabling the generator to learn synthesis strategies that reduce the detector confidence while preserving speech quality.

\subsection{Problem Formulation}

The speech deepfake detection problem can be formally defined as follows. 
Let $\mathbf{x} \in \mathbb{R}^T$ denote a discrete-time speech signal with class label $y \in \{0,1\}$, where $y=0$ indicates an authentic signal and $y=1$ a synthetically generated one.
The goal of this task is to develop a speech deepfake detector $\mathcal{D}: \mathbb{R}^T \rightarrow [0,1]$ able to estimate the class of $\mathbf{x}$ as $\hat{y} = \mathcal{D}(\mathbf{x})$, where $\hat{y}$ represents the likelihood of $\mathbf{x}$ being synthetic.

The goal of an adversarial attack is to generate a signal $\tilde{\mathbf{x}}$ that is perceptually similar to $\mathbf{x}$ but induces an incorrect prediction by the target detector $\mathcal{D}$, such as
\begin{equation}
   \mathcal{D}(\tilde{\mathbf{x}}) < \mathcal{D}(\mathbf{x}) \quad \text{subject to} \quad \mathbf{x} \approx \tilde{\mathbf{x}},
\end{equation}
where $\approx$ indicates perceptual similarity.
% \begin{equation}
%     \mathcal{D}(\tilde{\mathbf{x}}) < \mathcal{D}(\mathbf{x}) \quad \text{subject to} \quad \|\mathbf{x} - \tilde{\mathbf{x}}\|_p \leq \epsilon,
% \end{equation}
% where $\|\cdot\|_p$ represents the $\ell_p$-norm distortion metric, and $\epsilon$ constrains permissible signal modification.

\subsection{Adversarial TTS Architecture}

Let us define a \gls{tts} generator $\mathcal{G}$ as a system that receives a text string $\mathbf{t}$ as input and returns a speech signal $\mathbf{x}$ as output, as in $\mathbf{x} = \mathcal{G}(\mathbf{t})$.
We consider the \gls{tts} generator $\mathcal{G}$ as the composition of two cascade modules:
\begin{itemize}[leftmargin=*]
    \item An \textit{Acoustic Model} $\mathcal{A}$, that maps text $\mathbf{t}$ to mel-spectrogram $\mathbf{M} \in \mathbb{R}^{F \times N}$ through
    \begin{equation}
        \mathbf{M} = \mathcal{A}(\mathbf{t}; \theta_{\mathcal{A}}),
    \end{equation} 
    where $\theta_{\mathcal{A}}$ denotes model parameters, $F$ the mel-frequency bins, and $N$ the temporal frames.
    \item A \textit{Vocoder} $\mathcal{V}$, which synthesizes $\mathbf{M}$ into a $T$-samples waveform $\mathbf{x} \in \mathbb{R}^T$ via
    \begin{equation}
        \mathbf{x} = \mathcal{V}(\mathbf{M}; \theta_{\mathcal{V}}).
    \end{equation}
\end{itemize}
The resultant speech sample is defined as $\mathbf{x} = \mathcal{V}(\mathcal{A}(\mathbf{t}))$.

Since $\mathbf{x}$ is synthetically generated, its class label is $y = 1$, and an ideal detector should assign it a high $\hat{y}$ value, meaning a high probability of belonging to the synthetic class.
Our objective is therefore to adapt $\mathcal{G}$ so that the synthesized speech $\tilde{\mathbf{x}}$ is instead classified as authentic.
Rather than retraining the entire \gls{tts} system, we keep the acoustic model $\mathcal{A}$ fixed and fine-tune only the vocoder $\mathcal{V}$. This choice preserves the linguistic and speaker representations learned by the acoustic model while allowing the waveform generation process to suppress detector-relevant forensic cues.

To do so, we optimize the vocoder parameters $\theta_{\mathcal{V}}$ using a composite loss that combines the original vocoder training objective $\mathcal{L}_{\text{V}}$ with an adversarial detection loss $\mathcal{L}_{\text{D}}$. 
The original vocoder loss preserves the synthesis quality learned during standard training, whereas the adversarial term encourages the generated waveform to be classified as authentic by the target detector. Importantly, during this optimization, the parameters of the detector $\mathcal{D}$ remain frozen and gradients are back-propagated only through the detector to update the vocoder.
The overall loss is therefore defined as:
\begin{equation}
    \mathcal{L}_{\text{total}} = \mathcal{L}_{\text{V}} + \lambda \mathcal{L}_{\text{D}},
    \label{eq:det_loss}
\end{equation}
where $\lambda$ is a weighting factor.
The detection loss $\mathcal{L}_{\text{D}}$ is computed by evaluating the synthesized waveform with the target detector and encouraging the detector to predict the authentic class. Formally, it is defined as:
\begin{equation}
    \mathcal{L}_{\text{D}} = \mathbb{E}_{\mathbf{t}} \left[ -\log\left(1 - \mathcal{D}\left( \mathcal{V}(\mathcal{A}(\mathbf{t})) \right) \right) \right].
\label{eq:det_loss2}
\end{equation}
This corresponds to minimizing the binary cross-entropy with respect to the authentic class. Consequently, the gradients propagated through the detector encourage the vocoder to generate waveforms whose forensic characteristics resemble those of genuine speech, while the original vocoder objective prevents the optimization from degrading synthesis quality.

The proposed formulation naturally extends to multiple target detectors. Instead of optimizing against a single detector, $\mathcal{L}_{\mathcal{D}}$ can be expressed as a weighted combination of detector-specific losses:
\begin{equation}
    \mathcal{L}_{\text{D}} = \sum_{k=1}^K \alpha_k \mathbb{E}_{\mathbf{t}} \left[ -\log\left(1 - \mathcal{D}_k\left( \mathcal{V}(\mathcal{A}(\mathbf{t})) \right) \right) \right],
\end{equation}
where $\mathcal{D}_k$ denotes the $k$-th detector and $\alpha_k$ weights detector-specific objectives.
By minimizing the composite objective, the vocoder jointly optimizes speech fidelity and adversarial effectiveness, producing synthetic speech that preserves perceptual quality while reducing the confidence of one or more target speech deepfake detectors. 
Figure~\ref{fig:tts-pipeline} shows the complete adversarial \gls{tts} pipeline.

\section{Experimental setup}
\label{sec:setup}

In this section, we describe the experimental setup adopted to evaluate the proposed framework. 
We first present the considered synthesis and detection systems, followed by the training procedure and the datasets used throughout the experiments.

\subsection{Considered Systems}

The proposed framework is built upon a two-stage \gls{tts} pipeline composed of an acoustic model and a neural vocoder. As acoustic model $\mathcal{A}$, we employ an enhanced version of FastSpeech2~\cite{ren2020fastspeech}, incorporating the modifications proposed in~\cite{chen2021adaspeech, bredin2020pyannote, binkowski2019high} to improve speaker conditioning and synthesis quality.
Specifically, the model integrates Utterance- and Phoneme-Level Encoders, pyannote-based speaker embeddings~\cite{bredin2020pyannote}, a Conditional Layer for speaker adaptation, and \glspl{rwd}~\cite{binkowski2019high}. We refer the reader to the corresponding references for a detailed description of these components. 

As vocoder $\mathcal{V}$, we adopt BigVGAN~\cite{lee2022bigvgan}, a state-of-the-art neural vocoder for waveform synthesis. The acoustic model remains fixed throughout our experiments, while only the vocoder is fine-tuned using the detector-guided objective described in Section~\ref{sec:method}. This design isolates the effect of adversarial optimization on waveform generation while preserving the linguistic and speaker representations learned by the acoustic model.

To assess the impact of detector-guided optimization, we train three different vocoder variants:
\begin{itemize}[leftmargin=*]
   \item $\mathcal{V}$: The original vocoder, trained only with the original loss function $\mathcal{L}_\text{V}$ proposed in~\cite{lee2022bigvgan}, without any adversarial loss.
   \item $\mathcal{V^\text{R}}$: Trained including $\mathcal{L}_\text{D}$ with $K=1$, to fool a single detector (RawNet2).
   \item $\mathcal{V^\text{RT}}$: Trained with a multi-detector loss $\mathcal{L}_\text{D}$, aiming to deceive the $K=2$ detectors together (RawNet2 and TSSD).
\end{itemize}

Since detector-guided optimization is performed exclusively during offline training, the computational cost is incurred only once during fine-tuning. At inference time, the proposed framework has the same computational complexity and latency as the original BigVGAN vocoder, as no additional detector inference or waveform post-processing is required.

Finally, we evaluate our adversarial framework against three end-to-end speech deepfake detectors $\mathcal{D}$ that operate directly on raw audio waveforms: RawNet2~\cite{tak2021end}, TSSD~\cite{hua2021towards}, and AASIST~\cite{jung2022aasist}.
RawNet2 and TSSD are used both during adversarial optimization and evaluation, representing the white-box attack scenario. In contrast, AASIST is never involved during training and is used exclusively during testing to assess the transferability of the proposed attack under black-box conditions. 
The three vocoder configurations considered throughout the paper are summarized in Table~\ref{tab:vocoder_configs}.

\begin{table}
\caption{Vocoder configurations considered throughout the experiments.}
\label{tab:vocoder_configs}
\centering
\begin{tabular}{ccc}
\toprule
Name & Detector-guided training & Purpose \\
\midrule
$\mathcal{V}$           & None                     & Baseline synthesis \\
$\mathcal{V^\text{R}}$  & RawNet2                  & Single-detector attack \\
$\mathcal{V^\text{RT}}$ & RawNet2 + TSSD           & Multi-detector attack \\
\bottomrule
\end{tabular}
\end{table}

\subsection{Training parameters}

The acoustic model was trained following the original FastSpeech2 implementation, adopting the optimization strategy described in~\cite{ren2020fastspeech}. Likewise, the baseline BigVGAN vocoder was trained using the configuration proposed in~\cite{lee2022bigvgan}. Unless otherwise specified, we retain the original training hyperparameters of both architectures. 

The adversarial fine-tuning stage differs only in the optimization objective.
Starting from the baseline vocoder, we optimize the parameters according to the composite loss introduced in Eq.~\ref{eq:det_loss}, combining the original vocoder objective with the detector-guided loss. During this phase, the acoustic model and the detector networks remain frozen, and gradients are back-propagated only to update the vocoder parameters.

In practice, Eq.~\ref{eq:det_loss2} is implemented as the binary cross-entropy loss between the detector prediction and the authentic class, as in
\begin{equation}
	\mathcal{L}_\text{D} = \lambda \cdot \text{CE}(\text{$\mathcal{D}(\mathbf{x})$}, \text{0}) ,
\end{equation}
where the target label $0$ denotes authentic speech and $\lambda = 0.05$ is a scaling constant that controls the contribution of the adversarial objective. Minimizing this loss encourages the vocoder to generate speech that is increasingly difficult for the target detector to classify as synthetic. 

All detectors were trained under identical conditions using the Adam optimizer with cross-entropy loss. Additional implementation details, including optimizer parameters and learning-rate schedules, follow the original implementations of the corresponding models. All experiments were conducted on a workstation equipped with an Intel i9-14900K CPU, 64 GB RAM, and an NVIDIA RTX 4090 GPU.
The code used for training and evaluating the models, along with audio samples generated by the proposed \gls{tts} systems, is accessible at this link\footnote{\url{https://github.com/polimi-ispl/adversarial-tts/}}.

\subsection{Datasets}

We trained the acoustic model $\mathcal{A}$ and the vocoder $\mathcal{V}$ using LibriTTS~\cite{zen2019libritts}, a dataset derived from LibriSpeech~\cite{panayotov2015librispeech} and specifically processed for \gls{tts} applications.
This provides approximately $260$ hours of high-quality speech from $1158$ speakers sampled at $f_\text{s}=$~\SI{22.05}{\kilo\hertz}. Following the preprocessing adopted in our experiments, utterances shorter than \SI{3}{\second} and recordings flagged for quality issues were discarded. The resulting dataset was randomly split into \num{90}\% for training and \num{10}\% for validation.

To fine-tune BigVGAN, we considered both real and synthetic inputs, utilizing mel-spectrograms extracted from LibriTTS signals, as well as those generated by FastSpeech2.
This approach adapts the vocoder to the specific characteristics of the output of the acoustic model, resulting in a more stable and coherent speech generation.

We trained the deepfake detectors $\mathcal{D}$ using an internal dataset balanced between real and synthetic speech for a total of \num{251154} tracks sampled at $f_\text{s}=$~\SI{22.05}{\kilo\hertz}. Authentic recordings were drawn from LibriTTS, whereas synthetic speech was generated using the baseline \gls{tts} pipeline employing the vocoder $\mathcal{V}$.

To evaluate the effectiveness of detector-guided vocoder training, we constructed three test datasets that share the same authentic speech recordings while differing only in the synthetic samples. Specifically, the fake speech in each dataset was generated using one of the three vocoder variants ($\mathcal{V}$, $\mathcal{V}^\text{R}$, and $\mathcal{V}^\text{RT}$). This controlled protocol isolates the effect of adversarial fine-tuning by ensuring that the authentic data remain identical across all evaluation scenarios.

\begin{figure}
\centering
\includegraphics[width=.75\columnwidth]{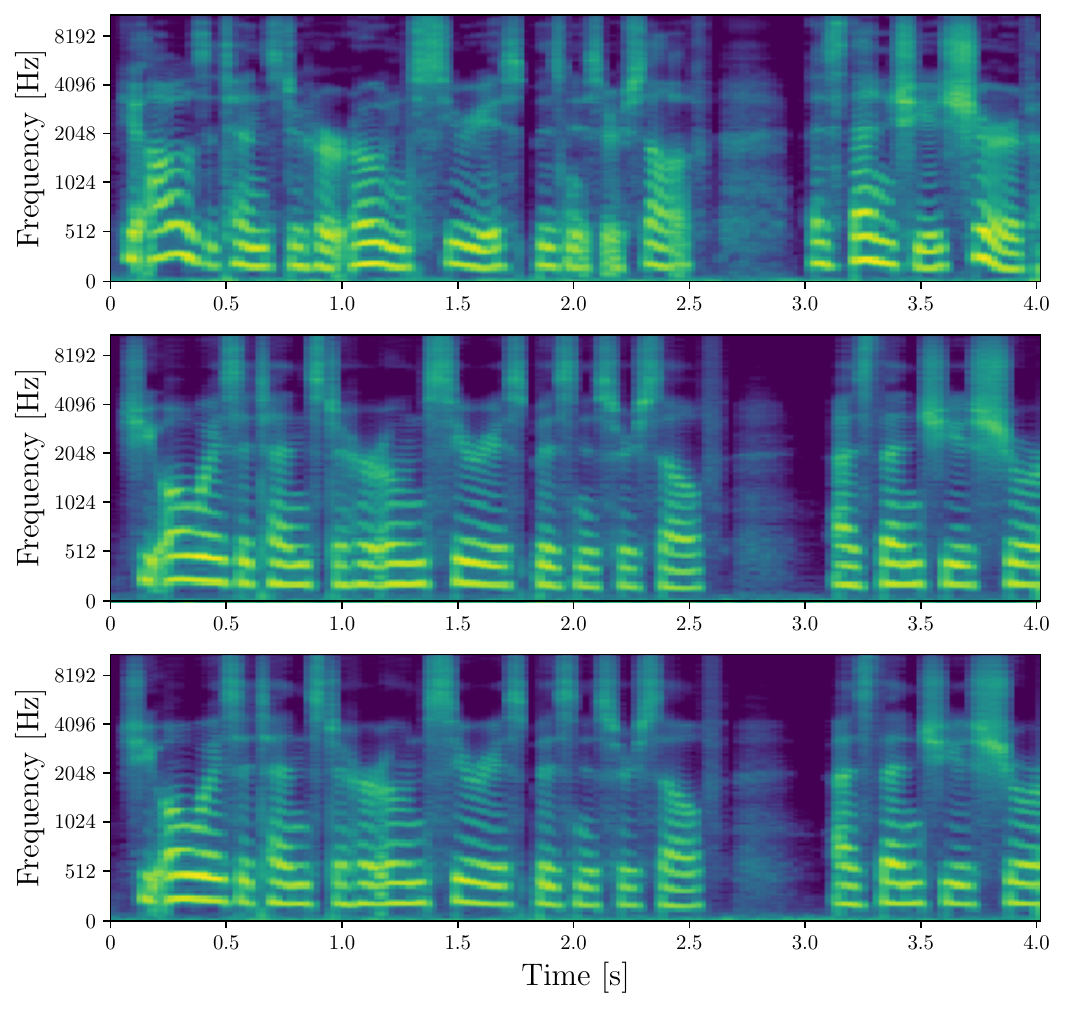}
\caption{Mel-spectrograms of real and synthetic signals. The real signal (top) is from LibriTTS, the synthetic ones are from $\mathcal{V}$ (middle) and $\mathcal{V}^\text{R}$ (bottom).}
\label{fig:spectrograms}
%\vspace{-.75em}
\end{figure}

Figure~\ref{fig:spectrograms} presents the mel-spectrograms of a real test track alongside its synthetic counterparts from $\mathcal{V}$, and $\mathcal{V}^\text{R}$, showing that the differences between them are almost imperceptible.

\begin{table*}[t]
\centering
\caption{Detection performance of the considered speech deepfake detectors $\mathcal{D}$ against different iterations of the vocoder $\mathcal{V}$.}
\label{tab:results}
\begin{tabular}{ccccccccccccc}
\toprule
\multirow{2.5}{*}{} & & \multicolumn{3}{c}{RawNet2}  & & \multicolumn{3}{c}{TSSD} & & \multicolumn{3}{c}{AASIST} \\ \cmidrule(lr){3-5} \cmidrule(lr){7-9} \cmidrule(lr){11-13}
         & & EER      & Bal. Acc. & AUC      & & EER      & Bal. Acc. & AUC & & EER      & Bal. Acc. & AUC      \\ \midrule
% $\mathcal{V}^\text{PRE}$      & & 70.11\% & 29.89\% & 73.16\% & & 63.42\% & 36.58\% & 70.15\% \\ 
$\mathcal{V}$    & & 5.8\% & 94.15\% & 97.08\% & & 3.40\% & 95.73\% & 98.80\% & & 0.85\% & 99.16\% & 99.95\% \\ 
$\mathcal{V}^\text{R}$    & & 39.05\% & 60.85\%  & 64.58\% & & 38.33\% & 60.54\% & 66.08\%  & & 21.09\% & 78.91\% & 87.07\% \\ 
$\mathcal{V}^\text{RT}$ & & 43.87\% & 55.17\%  & 57.72\% & & 43.31\% & 57.03\% & 59.74\%  & & 23.26\% & 76.75\% & 84.55\% \\ \bottomrule
\end{tabular}
% \vspace{-.75em}
\end{table*}

\section{Results}
\label{sec:results}

In this section, we evaluate the proposed framework from three complementary perspectives. First, we assess its ability to degrade speech deepfake detectors under both white-box and black-box conditions. Second, we investigate whether jointly optimizing against multiple detectors improves attack transferability. Finally, we evaluate the perceptual quality of the generated speech using both objective and subjective metrics.

Detection performance is measured in terms of \gls{auc}, Balanced Accuracy, and \gls{eer}. 
While ideal detector performance corresponds to high \gls{auc} and Balanced Accuracy and low \gls{eer}, the objective of the proposed adversarial framework is the opposite: to reduce the detector’s ability to discriminate between authentic and synthetic speech. Since only the synthetic signals are modified while the authentic ones remain unchanged, the theoretical optimum of a successful attack corresponds to 50\% for all three metrics, indicating random discrimination between the two classes.

We evaluate the three speech deepfake detectors (RawNet2, TSSD, and AASIST) using speech synthesized by the three vocoder configurations introduced in Section~\ref{sec:setup}, namely the baseline vocoder $\mathcal{V}$, the single-detector adversarial vocoder $\mathcal{V}^\text{R}$, and the multi-detector adversarial vocoder $\mathcal{V}^\text{RT}$. This experimental protocol enables us to separately evaluate baseline detector performance, white-box attack effectiveness, and transferability to previously unseen detector architectures.

Table~\ref{tab:results} summarizes the detection performance obtained under the three evaluation settings. As expected, all detectors achieve excellent performance on speech generated by the baseline vocoder $\mathcal{V}$, with \gls{auc} values above 97\% and \gls{eer} below 6\%. Since these synthetic samples closely match those used during detector training, this experiment establishes a strong reference point for assessing the effectiveness of the proposed adversarial optimization. Among the evaluated systems, AASIST achieves the highest detection accuracy, although all three detectors reliably distinguish authentic from synthetic speech.

Detector-guided fine-tuning substantially degrades the performance of the targeted detectors. When optimizing exclusively against RawNet2, its \gls{eer} increases from $5.8\%$ to $39.1\%$, approaching the theoretical upper bound of $50\%$ associated with random guessing. Interestingly, TSSD also experiences a comparable degradation (\gls{eer}$ = 38.3\%$) despite not being explicitly optimized during training, suggesting that the learned adversarial modifications suppress forensic cues that are not entirely detector-specific. Even the unseen AASIST detector exhibits a marked reduction in performance, with its \gls{eer} increasing to $21.1\%$, indicating that the proposed optimization already provides a significant degree of transferability beyond the targeted architecture.

Joint optimization against RawNet2 and TSSD further improves the attack. Both targeted detectors exhibit \gls{eer} values above $43\%$, while their Balanced Accuracy decreases to approximately $56\%$, indicating detector performance close to random discrimination. More importantly, the performance of AASIST also decreases further, reaching an \gls{eer} of $23.3\%$ despite never being involved during adversarial optimization. This result indicates that jointly optimizing against multiple detectors improves attack transferability by encouraging the vocoder to suppress forensic cues shared across different detector architectures rather than overfitting to a single target model.

\begin{figure}
\centering
\includegraphics[width=.55\columnwidth]{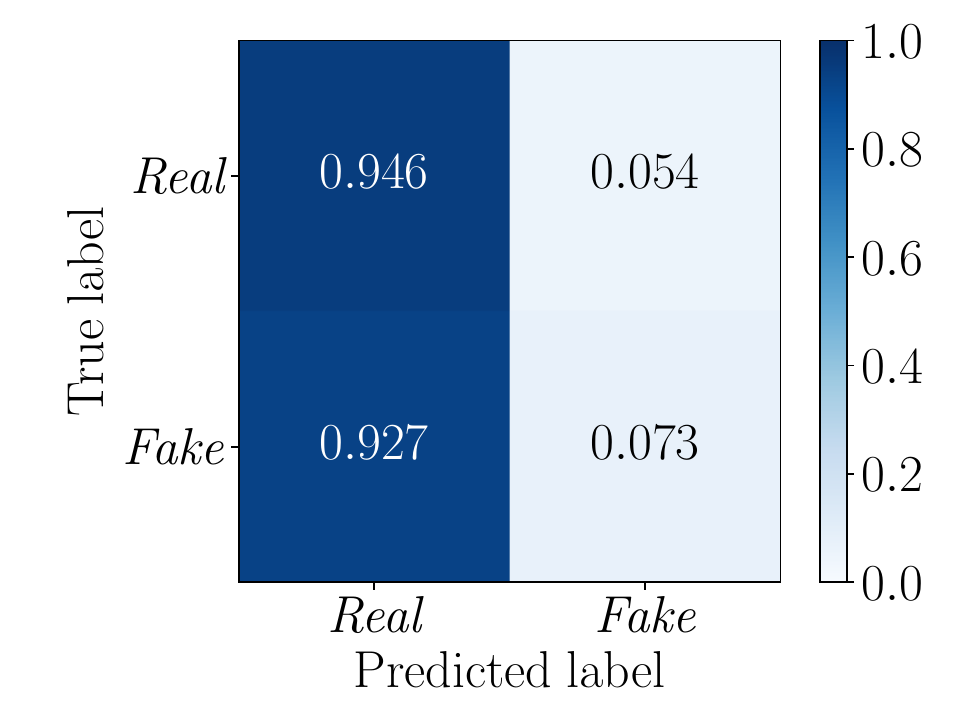}
\caption{Confusion matrix obtained by testing RawNet2 against the test dataset containing fake tracks generated by the vocoder $\mathcal{V}^\text{RT}$.}
\label{fig:cmatrix}
\end{figure}

Figure~\ref{fig:cmatrix} shows the detection performance of RawNet2 when tested against data generated by $\mathcal{V}^\text{RT}$, via a confusion matrix.
Almost all the synthetic tracks are misclassified as real, with an exceptionally high \gls{tnr} (\num{0.946}) and low \gls{tpr} (\num{0.073}), aligning with our adversarial goal.
With a balanced accuracy of \num{55}\%, the proposed attack effectively reduces the detector's performance to near-random guessing.

Beyond evaluating adversarial effectiveness, we also investigate the perceptual quality of the generated speech. We hypothesize that detector-guided optimization not only suppresses detector-relevant forensic traces but also reduces synthesis artifacts that distinguish generated speech from authentic recordings. To test this hypothesis, we first compute the \gls{fad}~\cite{kilgour2019fr} between authentic speech and the three synthetic datasets using CLAP~\cite{elizalde2024natural} embeddings. The \gls{fad} decreases consistently from $31.25$ for the baseline vocoder $\mathcal{V}$ to $22.95$ for $\mathcal{V}^\text{R}$ and $17.81$ for $\mathcal{V}^\text{RT}$, corresponding to an improvement of approximately $25\%$ at each optimization stage. This progressive reduction suggests that adversarial fine-tuning moves the distribution of synthesized speech closer to that of authentic recordings while simultaneously improving resistance to forensic detection. 

\begin{figure}
\centering
\includegraphics[width=.85\columnwidth]{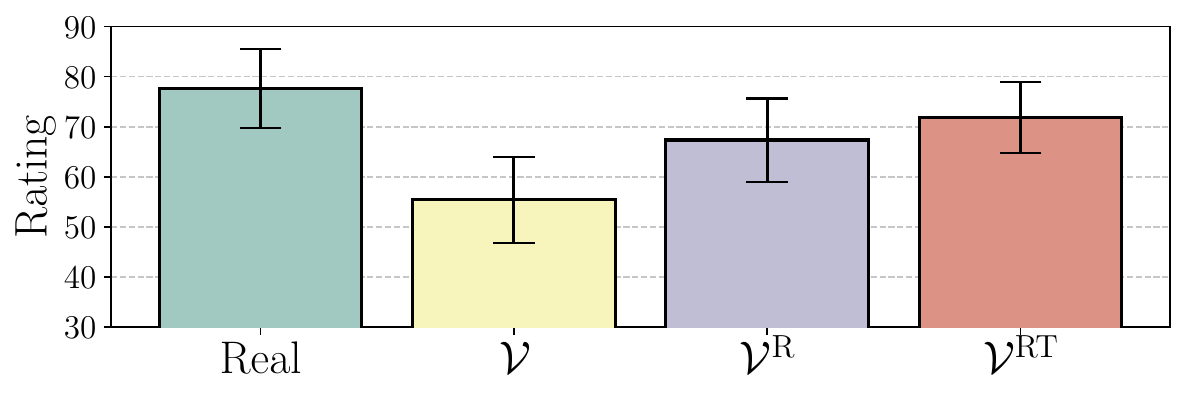}
\caption{Results of the perceptual listening test evaluating the realism of the real and synthetic tracks generated with different iterations of the vocoder $\mathcal{V}$.}
\label{fig:mushra}
% \vspace{-1em}
\end{figure}

To complement the objective evaluation, we conducted a MUSHRA-like perceptual listening test with \num{20} expert listeners to assess the level of realism of the generated tracks.
The study was anonymous and GDPR-compliant, with informed consent obtained and no personal data collected.
Figure~\ref{fig:mushra} presents the results of this analysis, indicating that the fine-tuned vocoders $\mathcal{V}^\text{R}$ and $\mathcal{V}^\text{RT}$ were perceived as more realistic compared to the initial $\mathcal{V}$ vocoder, progressively approaching the quality of real audio.
Together with the FAD analysis, these results indicate that our adversarial fine-tuning not only enhances attack robustness but also increases the perceptual quality of synthesized speech, making it a promising strategy for advancing adversarial \gls{tts} systems.

The observed improvement in perceptual quality is particularly interesting because the detector loss is never optimized explicitly for human perception. A possible explanation is that speech deepfake detectors rely on subtle synthesis artifacts that are also perceptually undesirable. By encouraging the vocoder to suppress these detector-relevant cues while preserving the original synthesis objective, the optimization may simultaneously improve detector evasion and speech naturalness. Although further investigation is required to validate this hypothesis, the consistent improvements observed in both FAD and the listening test suggest a strong relationship between forensic artifacts and perceptual quality.

Overall, the experiments demonstrate three main findings. First, detector-guided vocoder optimization substantially degrades the performance of targeted speech deepfake detectors under white-box conditions. Second, jointly optimizing against multiple detectors improves transferability to previously unseen detector architectures. Finally, adversarial fine-tuning unexpectedly improves both objective and subjective measures of speech quality, indicating that suppressing detector-relevant forensic cues can also increase the perceptual realism of synthesized speech.

% We acknowledge that these results present dual-use implications.
% While our unique goal is to expose vulnerabilities in forensic deepfake detectors, thereby contributing to the development of more robust countermeasures, the proposed adversarial framework could be misused to create synthetic signals that evade detection.
% We recognize that this lowers the barrier for generating deceptive content. However, we firmly believe that investigating these vulnerabilities is a necessary step in the ongoing arms race between attackers and defenders.
% To this end, we encourage the use of this work solely for defensive security research and advocate for the continued development of ethical guidelines and governance structures to mitigate the potential misuse of generative audio technology.

\section{Conclusions}
\label{sec:conclusions}

In this paper, we presented a detector-guided \gls{tts} framework capable of generating speech that is intrinsically adversarial to speech deepfake detectors. By integrating one or more frozen detector models into the vocoder optimization process, the proposed approach directly learns waveform generation strategies that reduce detector confidence while preserving the perceptual quality of the synthesized speech. Unlike existing adversarial attacks that rely on post-processing perturbations, the proposed framework produces adversarial speech during synthesis, eliminating the need for any additional processing at inference time.

Extensive experiments against three state-of-the-art speech deepfake detectors showed the effectiveness of the proposed approach under both white-box and black-box conditions. Detector-guided optimization substantially degraded the performance of the targeted detectors, while joint optimization against multiple detectors further improved transferability to previously unseen architectures. Interestingly, objective and subjective evaluations consistently showed that adversarial fine-tuning also enhances the perceptual realism of the generated speech, suggesting that suppressing detector-relevant forensic cues may simultaneously reduce synthesis artifacts.

The proposed framework also highlights the growing challenge of securing speech synthesis systems against increasingly sophisticated adversarial attacks. Although our objective is to expose weaknesses in current forensic methods and support the development of more robust countermeasures, the same techniques could potentially be misused to generate synthetic speech that is more difficult to detect. We therefore advocate the use of this work exclusively for defensive security research and encourage continued investigation into robust detection methods, benchmarking protocols, and ethical guidelines for generative audio technologies. 

While the proposed framework demonstrates strong performance against three representative speech deepfake detectors, the evaluation is limited to a single \gls{tts} architecture and a controlled synthesis scenario.
Future work will investigate additional vocoders, multilingual datasets, real-world transmission conditions, and more computationally efficient optimization strategies, with the ultimate goal of developing speech deepfake detectors that remain robust against adversarially trained synthesis models.

% \newpage

\bibliographystyle{IEEEtran}
\bibliography{bibliography}

\end{document}